\documentclass[epj]{webofc}
\usepackage[utf8]{inputenc}
\usepackage[varg]{txfonts}   
\usepackage{booktabs}
\usepackage{xcolor}
\definecolor{darkred}{rgb}{0.4,0.0,0.0}
\definecolor{darkgreen}{rgb}{0.0,0.4,0.0}
\definecolor{darkblue}{rgb}{0.0,0.0,0.4}
\usepackage[bookmarks,linktocpage,colorlinks,
    linkcolor = darkred,
    urlcolor  = darkblue,
    citecolor = darkgreen]{hyperref}
\usepackage{subfigure}
\usepackage{bm}
\wocname{EPJ Web of Conferences}
\woctitle{Lattice2017}
\newcommand{\Bs}{{\textnormal{B}_\textnormal{s}}}
\newcommand{\K}{\textnormal{K}}
\newcommand{\EB}{E}
\newcommand{\C}{\mathcal{C}}
\newcommand{\CB}{\C^\Bs}
\newcommand{\CBK}{\C^\textnormal{3pt}} 

\newcommand{\ord}[1]{\mathcal{O}\left(#1\right)}

\newcommand{\tk}{{t_\K}}
\newcommand{\tb}{{t_\Bs}}

\newcommand{\stat}{\textnormal{stat}}
\newcommand{\kin}{\textnormal{kin}}
\newcommand{\spin}{\textnormal{spin}}
\newcommand{\jn}{\textnormal{j}}
\newcommand{\kn}{\textnormal{k}}
\newcommand{\mh}{m_\textnormal{b}}

\newcommand{\Ri}{\mathcal{R}^{\textnormal{I}}}
\newcommand{\Rii}{\mathcal{R}^{\textnormal{II}}}
\newcommand{\dri}{\rho^\textnormal{I}}
\newcommand{\drii}{\rho^\textnormal{II}}

\newcommand{\eq}[1]{Eq.~(\ref{#1})}

\begin{document}
\selectlanguage{english}
\title{HQET form factors for $\Bs\to\K\ell\nu$ decays beyond leading order }
\author{\firstname{Debasish} \lastname{Banerjee}\inst{1}\fnsep \and
\firstname{Mateusz} \lastname{Koren}\inst{1}\fnsep\thanks{Speaker,
\email{mateusz.koren@desy.de}} \and \firstname{Hubert} 
\lastname{Simma}\inst{1}\fnsep \and \firstname{Rainer} \lastname{Sommer}\inst{1}
}
\institute{John von Neumann Institute for Computing (NIC), DESY, Platanenallee
6, D-15738 Zeuthen, Germany }
\abstract{We compute semi-leptonic $\Bs$ decay form factors using Heavy Quark
Effective Theory on the lattice. To obtain good control of the $1/\mh$
expansion, one has to take into account not only the leading static order but
also the terms arising at $\ord{1/\mh}$: kinetic, spin and current insertions.
We show results for these terms calculated through the ratio method, using our
prior results for the static order. After combining them with
non-perturbative HQET parameters they can be continuum-extrapolated to give the
QCD form factor correct up to $\ord{1/\mh^2}$ corrections and without
$\ord{\alpha_s(\mh)^n}$ corrections.
}
\hfill DESY 17-172
\maketitle
\section{Introduction}\label{intro}

Weak decays of B-mesons play an important role in determining the parameters of
the Standard Model. In particular, the charmless charged-current semi-leptonic B
decays, such as $\Bs\to\K\ell\nu$, give a way to extract the poorly-constrained
$|V_\textnormal{ub}|$ element of the CKM matrix.

The semi-leptonic $\Bs\to\K\ell\nu$ decay is mediated by QCD matrix elements,
which in the rest frame of the $\Bs$ meson have the following form:
\begin{align}
(2m_\Bs)^{-1/2}\langle\K(\mathbf{p}_\K)|V^0(0)|\Bs(0)\rangle&=
h_\parallel(E_\K),\\
(2m_\Bs)^{-1/2}\langle\K(\mathbf{p}_\K)|V^k(0)|\Bs(0)\rangle&=p^k_\K
h_\perp(E_\K),
\end{align}
where $\mathbf{p}_\K$ is the Kaon momentum and the vector current is
defined as $V^\mu(x) = \overline\psi_{\rm u}(x)\gamma^\mu\psi_{\rm b}(x)$.
Matrix elements $h_\parallel$ and $h_\perp$ are related to the often-used $f_+$,
$f_0$ by simple kinematic relations, cf.~Ref.~\cite{Bahr:2016ayy}.

Due to its large mass, the b quark requires special treatment to be discretized
on the lattice without large cutoff effects. Our approach is to use Heavy Quark
Effective Theory (HQET) \cite{Sommer:2015hea}. Results at the leading order of
HQET, the static approximation, are described in \cite{Bahr:2016ayy}. There one
can estimate the systematic error of the truncation of higher orders to be of
order 15\% -- a number which is reduced to 1-2\% when adding the $\ord{1/\mh}$
terms. It is therefore important to include these terms to obtain
phenomenologically relevant results.

HQET expansion of a correlation function up to $\ord{1/\mh}$ is
\begin{equation}
\langle O\rangle = \langle O_\stat \rangle_\stat +
\omega_\kin {\textstyle \sum}_x \langle O_\stat
\mathcal{O}_\kin(x) \rangle_\stat + \omega_\spin
{\textstyle \sum}_x \langle O_\stat \mathcal{O}_\spin(x) \rangle_\stat +
{\textstyle \sum}_{\kn} \omega_{\kn}\langle O_{\kn}\rangle_\stat,
\end{equation}
where $\mathcal{O}_\kin(x) =
\overline\psi_\textnormal{b}(x)\mathbf{D}^2\psi_\textnormal{b}(x)$,
$\mathcal{O}_\spin(x) =
\overline\psi_\textnormal{b}(x)\,\bm{\sigma}\cdot\mathbf{B}\psi_\textnormal{b}(x)$
are the kinetic and spin terms, and $O_{\kn}$ correspond to additional operators
in the effective theory, which are discussed in Sec.~\ref{sec:curr}. The
(dimensionful) parameters of HQET $\omega_\spin, \omega_\kin, \omega_{\kn}
\propto1/\mh$  have to be determined by non-perturbative matching to QCD
\cite{Heitger:2003nj,DellaMorte:2013ega,Sommer:2015hea}. The notation
$\langle\cdot\rangle_\stat$ means that the expectation values are defined with
respect to the renormalizable static action.

For the computation of the matrix elements in the large volume we use
$N_\textnormal{f}=2$ CLS ensembles \cite{Fritzsch:2012wq}. Unless explicitly
stated otherwise, all the results presented here were obtained on ensemble N6,
with $a=0.048$~fm and $m_\pi=340$~MeV. The $\Bs$ meson is kept at rest, while
the Kaon has momentum $\mathbf{p}_\K = \frac{2\pi}{L}(1,0,0)$, which corresponds
to $|\mathbf{p}_\K|=0.535$~GeV and the momentum transfer
$q^2=21.23\;\textnormal{GeV}^2$. For more details on the ensembles used, see
Ref.~\cite{Bahr:2016ayy}. The heavy quark is discretized using HYP1 action
\cite{DellaMorte:2005yc} and the light quarks are smeared with several levels of
Gaussian smearing \cite{Gusken:1989ad,Alexandrou:1990dq,Bernardoni:2013xba}.

\section{Matrix elements at order $1/\mh$}\label{sec:1omdefs}

The HQET expansion up to order $1/\mh$ of the heavy-light two-point correlation
function is
\begin{equation}
\CB(t)  = \CB_\stat(t)
\left(1 + \omega_\jn \frac{\CB_\jn(t)}{\CB_\stat(t)}\right),
\end{equation}
where we schematically define\footnote{We suppress all the spacetime indices and
their corresponding sums, see Ref.~\cite{Bahr:2016ayy} for more explicit
definitions.} $\CB_\stat \equiv \langle P_\textnormal{bs} P_\textnormal{sb}
\rangle_\stat$ and $\CB_\jn \equiv \langle P_\textnormal{bs} P_\textnormal{sb}
\mathcal{O}_\jn \rangle_\stat$ with $P_{\textnormal{q}_1\textnormal{q}_2} =
\overline\psi_{\textnormal{q}_1}\gamma_5 \psi_{\textnormal{q}_2}$ and
$\jn\in\{\kin,\spin\}$.

The HQET expansion of the energy is $E^\Bs = E^\Bs_\stat + m_\textnormal{bare} +
\omega_\kin E^\Bs_\kin + \omega_\spin E^\Bs_\spin$ \cite{Bernardoni:2013xba} (in
the following we suppress the $\Bs$ index for readability). The contributions to
the energy at the leading and next-to-leading order can be extracted from the
large-time behaviour of the correlation functions:
\begin{align}
&-\partial_t \ln\CB_\stat(t) = \EB_\stat + \ord{e^{-\Delta Et}}, \\
&-\partial_t \frac{\CB_\jn(t)}{\CB_\stat(t)} = \EB_\jn + \ord{te^{-\Delta Et}}.
\end{align}

The three-point correlation functions, $\CBK_{\mu,\stat}\equiv\langle
P_\textnormal{su}V^\mu P_\textnormal{bs}\rangle_\stat$, can also be used to
obtain the energy contributions:
\begin{equation}
-\partial_\tb \frac{\CBK_{\mu,\jn}(\tk,\tb)}{\CBK_{\mu,\stat}(\tk,\tb)} =
\EB_\jn + \ord{\tb e^{-\Delta E\tb}}.
\end{equation}

From here on, for simplicity, we set that $\tk=\tb=t$ and drop the second
argument in the three-point functions and ratios.

After integrating the equations at order $1/\mh$ we obtain
\begin{align}
&\frac{\CB_\jn(t)}{\CB_\stat(t)} = A_\jn- \EB_\jn t +
\ord{te^{-\Delta Et}},\label{eq:rho2pt}\\
&\frac{\CBK_{\mu,\jn}(t)}{\CBK_{\mu,\stat}(t)} = A_{\mu,\jn}^\textnormal{3pt}
- \EB_\jn t + \ord{te^{-\Delta Et}},\label{eq:rho3pt}
\end{align}
where $A$ are the integration constants, which depend on the smearing
used (for notational clarity we keep the smearing indices implicit).

We define two ratios from which we can obtain the desired bare matrix elements:
\begin{align}
  \Ri_\mu(t) = &
    \frac{\CBK_\mu(t)}{\big[\C^\K(2t)\CB(2t)\big]^{1/2}},\label{eq:ri}\\[3pt]
  \Rii_\mu(t) = &
    \frac{\CBK_\mu(t)}{\big[\C^\K(t)\CB(t)\big]^{1/2}}
    e^{(\bar{E}^\K+\bar{E}^\Bs)\frac{t}{2}},\label{eq:rii}
\end{align}
where for precision we set the effective energies $\bar{E}^\K,\bar{E}^\Bs$ to
their ground-state values as extracted from the plateaux and GEVP plateaux
respectively. This reduces the statistical error compared to the time-dependent
effective energies at the expense of any systematic error in the determination
of the energies propagating into the ratios. However, the ground-state energies
are well under control.

Let us start with ratio $\Ri$ which is simpler theoretically, because it
requires no extra input apart from the correlation functions, but has larger
statistical errors, due to the use of the heavy-light two-point function at time
separation $2t$. 

In the following, we use the symbol $\cong$ for relations which hold in the
asymptotic large-time limit and up to terms of $\ord{1/\mh^2}$. HQET expansion
of the ratio is
\begin{equation}
\Ri_\mu(t) \cong
\Ri_{\mu,\stat}(t)\left[1
+ \omega_\jn\left(\frac{\CBK_{\mu,\jn}(t)}{\CBK_{\mu,\stat}(t)} -
\frac12\frac{\CB_\jn(2t)}{\CB_\stat(2t)}\right) +
\omega_{\kn}\frac{\CBK_{\mu,\kn}(t)}{\CBK_{\mu,\stat}(t)}\right]
\equiv
\Ri_{\mu,\stat}(t)\left[1
+ \omega_\jn\dri_{\mu,\jn} +
\omega_{\kn}\rho_{\mu,\kn}\right],
\end{equation}
where $\jn$ is implicitly summed over $\{\kin,\spin\}$ and $\kn$ over the
additional vector-current contributions, see Sec.~\ref{sec:curr}. At
large times $\rho$-s correspond to the $1/\mh$ corrections to the bare HQET
matrix elements.

The same procedure can be applied to $\Rii$:
\begin{align}
\Rii_\mu(t)
\cong &\,
\Rii_{\mu,\stat}(t)\left[1+\omega_\jn\left(\frac{\CBK_{\mu,\jn}(t)}
{\CBK_{\mu,\stat}(t)} - \frac12
\frac{\CB_\jn(t)}{\CB_\stat(t)}+\frac{\EB_\jn t}{2}\right) +
\omega_{\kn}\rho_{\mu,\kn}\right]
\label{eq:riiek}\\
\cong &\,
\Rii_{\mu,\stat}(t)\left[1+\omega_\jn\left(\frac{\CBK_{\mu,\jn}(t)}
{\CBK_{\mu,\stat}(t)} -
\frac{\CB_\jn(t)}{\CB_\stat(t)}
+ \frac{A_\jn}{2}\right) +
\omega_{\kn}\rho_{\mu,\kn}\right],\label{eq:riirho}
\end{align}
where \eq{eq:riiek} and \eq{eq:riirho} are related by \eq{eq:rho2pt}.
Analogously to the previous case we label the coefficients multiplying
$\omega_\jn$ as $\drii_{\mu,\jn}$. Note that both methods of calculating
$\drii_{\mu,\jn}$ require extra input (either $\EB_\jn$ or $A_\jn$) that must be
obtained through a fit, as described in the following subsection.

In addition, we note that there is one additional contribution to the matrix
elements at the 1/m level coming from an overall multiplicative renormalization
to the ratios $\Ri$ and $\Rii$, we refer the interested reader to
Refs.~\cite{Sommer:2015hea,Heitger:2003nj} for details.

\subsection{Results for the kinetic and spin insertions}
\label{sec:rat}

At large enough time, for each $\jn$ separately, both ratios must plateau at the
same value, from which we obtain the desired $1/\mh$ matrix-element
contribution. Using Eqs.~(\ref{eq:rho2pt})-(\ref{eq:rho3pt}), it can be also
written as
\begin{equation}
\dri_{\mu,\jn}(t) \cong \drii_{\mu,\jn}(t) \cong
A^\textnormal{3pt}_{\mu,\jn} - A_\jn/2.
\label{eq:plat}
\end{equation}
Note that, while both $A_\jn$, $A^\textnormal{3pt}_{\mu,\jn}$ depend on the
smearing, their combination in \eq{eq:plat} does not.

\begin{figure}[tbp]
  \centering \includegraphics[width=11cm,clip]{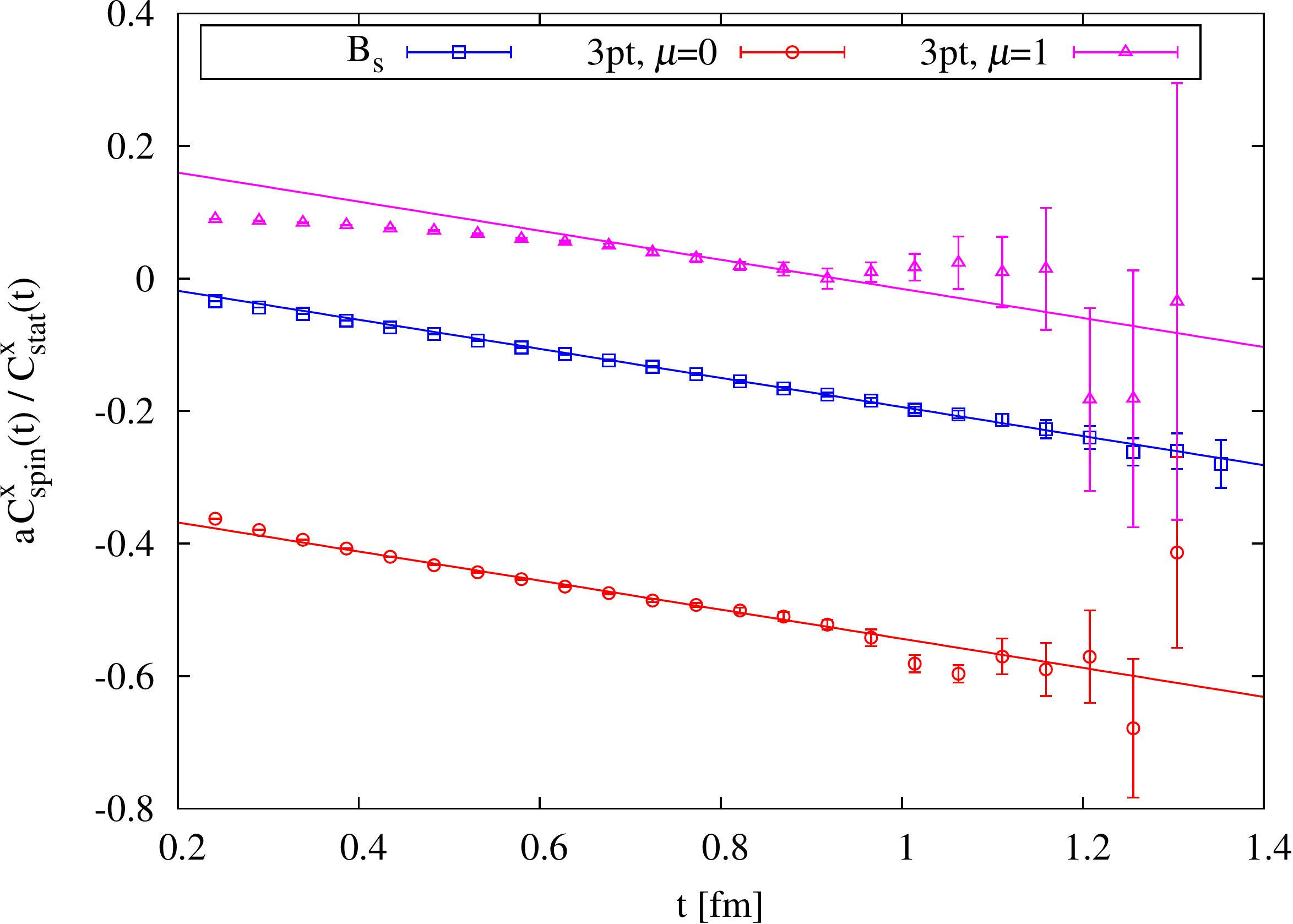}
  \caption{The example result of the simultaneous fit procedure for the spin
  insertions.}
  \label{fig:fit}
\end{figure}

We can extract $A_\jn$, $A^\textnormal{3pt}_{\mu,\jn}$ from fitting
Eqs.~(\ref{eq:rho2pt}) and (\ref{eq:rho3pt}). We choose to do a simultaneous fit
to both the equations, and both $\mu=0$ and 1. In this way we are utilizing the
fact that the linear slope (given by the energy contribution $\EB_\jn$) is
common in all of them. We find that this significantly improves the precision
that we can obtain, especially for the particularly demanding
$A^\textnormal{3pt}_{\mu=1,\jn}$. An illustration is given in
Fig.~\ref{fig:fit}. We use the fit range
$t\in[0.8~\textnormal{fm},1.4~\textnormal{fm}]$.

\begin{figure}[tbp]
  \centering
   \subfigure{\includegraphics[width=0.49\textwidth,clip]{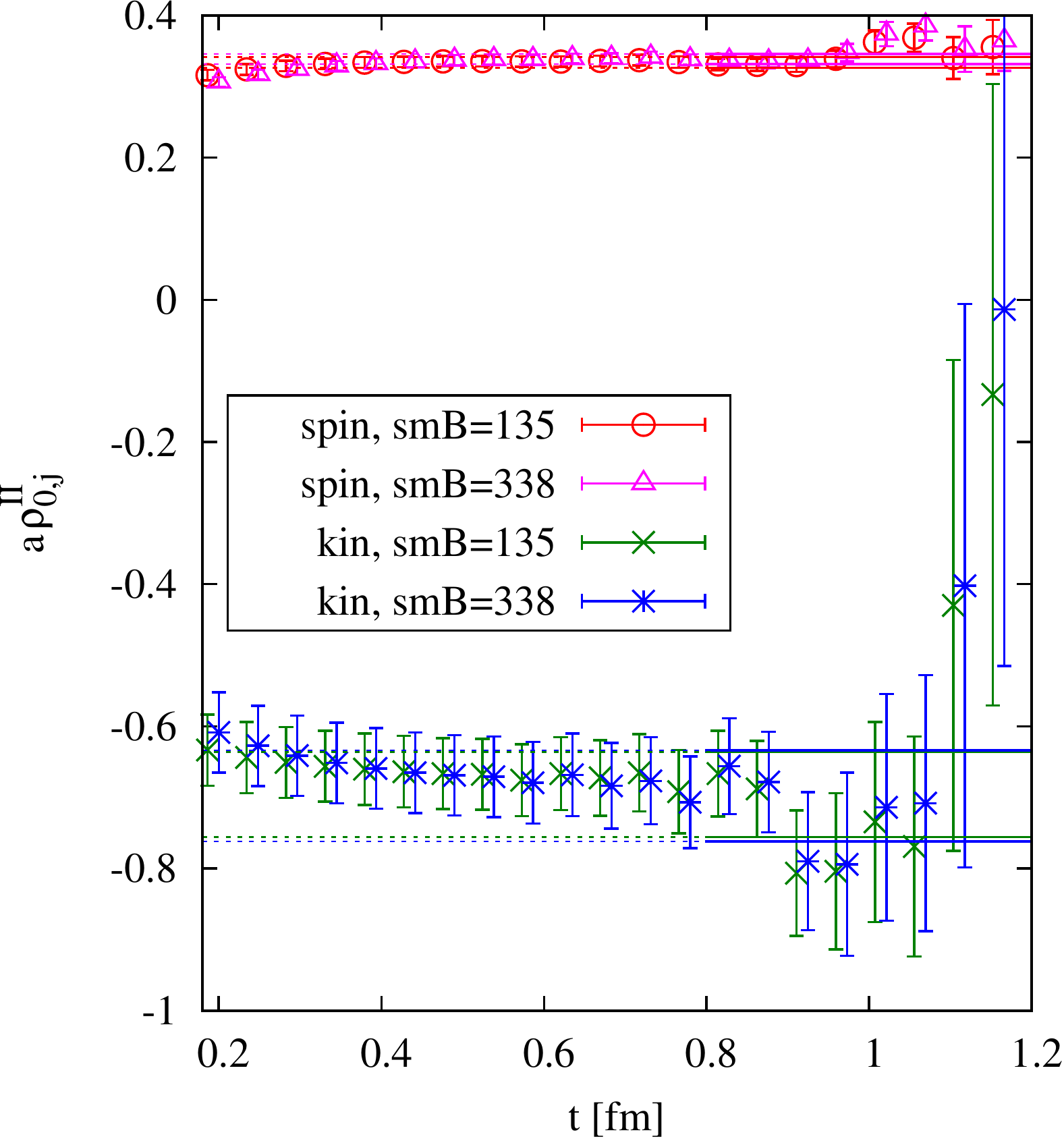}}
   \hfill
   \subfigure{\includegraphics[width=0.49\textwidth,clip]{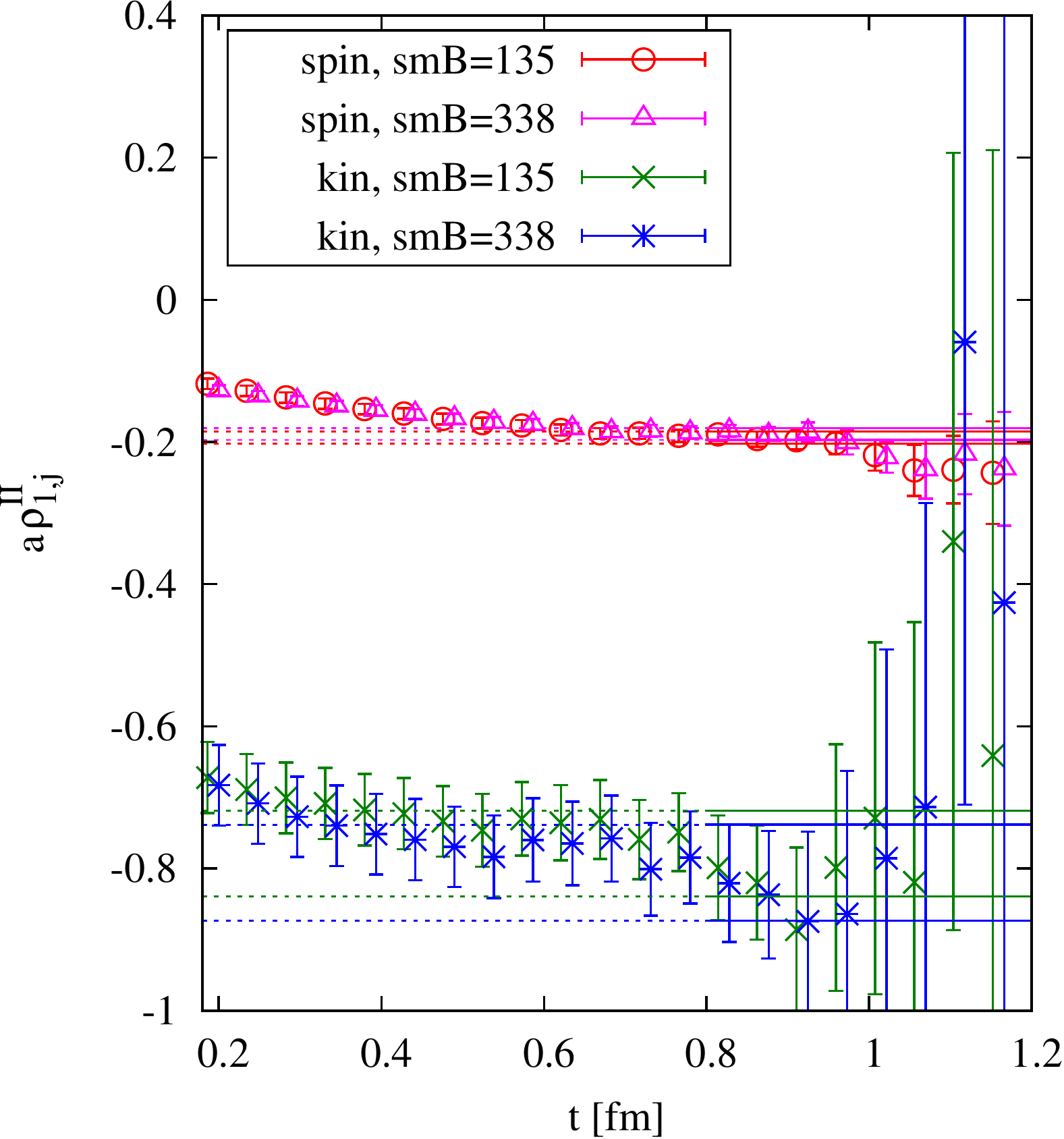}}
   \hfill
   \caption{Kinetic and spin insertions for $\mu=0$ (left) and $\mu=1$ (right)
   from the ratios and fits, for two different amounts of light-quark
   smearing. The data for different smearings are slightly shifted horizontally
   for greater visibility.}
  \label{fig:kinspin}
\end{figure}

In Fig.~\ref{fig:kinspin} we show the comparison of $\drii_{\mu,\jn}(t)$,
calculated using \eq{eq:riirho} with the $A_\jn$ from the simultaneous fit (one
can also calculate $A_\jn$ from the two-point functions only, the results are
perfectly consistent and have slightly larger errorbars). We also show the bands
coming from the fits. A good agreement between different light-quark smearings
is observed. In addition, the fit results for the highest smearing are collected
in Table~\ref{tab:kinspin}.

\begin{table}[tbp]
\centering
\begin{tabular}{cc}
 $\mu$ & $\drii_{\mu,\kin}$ 
 \\
 \hline
 0 & -0.70(6)  \\
 1 & -0.81(7)  \\
 \end{tabular}
 \quad\quad\quad
 \begin{tabular}{cc}
 $\mu$ & $\drii_{\mu,\spin}$ 
 \\
 \hline
 0 & 0.338(7) \\
 1 & -0.189(8)
 \end{tabular}
\caption{Fit results for the kin (left) and spin (right) contributions at the
highest light-quark smearing.}
\label{tab:kinspin}
\end{table}

The uncertainty of the kin contribution is dominant. Note, however, that the
previously determined subset of the HQET parameters \cite{Blossier:2012qu}
yields for $\omega_\spin$ a value which is approximately two times larger than
for $\omega_\kin$, therefore the overall contribution and uncertainty of the two
channels is more comparable.

\subsection{Vector current insertions}
\label{sec:curr}

\begin{table}[htbp!]
\begin{center}
 \begin{tabular}{ccccc}
 $\mu$ & $\kn$ & $V_{\mu,\kn}$ & $\omega^\textnormal{tree}_{\mu,\kn}\cdot\mh$ 
 & $\rho_{\mu,\kn}$
 \\
 \hline
 0 & 1 & $\overline\psi_\ell(x)\sum_l  \gamma_l
 \tfrac12(\nabla_l^\textnormal{S} -
 \overleftarrow\nabla_l^\textnormal{S})\psi_\textnormal{b}(x)$  & 1/2 &
 -0.0730(13)\\ 
 0 & 2 & $\overline\psi_\ell(x)\sum_l  \gamma_l
 \tfrac12(\nabla_l^\textnormal{S} +
 \overleftarrow\nabla_l^\textnormal{S})\psi_\textnormal{b}(x)$ & 1/2 &
  -0.0284$(\phantom{0}3)$ \\[2pt]
\hline 
 $i$ & 1 & $\overline\psi_\ell(x)\sum_l  \tfrac12 (\nabla_l^\textnormal{S} -
 \overleftarrow\nabla_l^\textnormal{S})\gamma_l\gamma_i\psi_\textnormal{b}(x)$ 
 & 1/2 & 0.3232(26)\\
 $i$ & 2 & $\overline\psi_\ell(x)\tfrac12 (\nabla_i^\textnormal{S} -
 \overleftarrow\nabla_i^\textnormal{S})\psi_\textnormal{b}(x)$ & -1 &
 0.0869(17) \\
 $i$ & 3 & $\overline\psi_\ell(x)\sum_l  \tfrac12 (\nabla_l^\textnormal{S} +
 \overleftarrow\nabla_l^\textnormal{S})\gamma_l\gamma_i\psi_\textnormal{b}(x)$ & 
 1/2 & 0.1083(12)\\
 $i$ & 4 & $\overline\psi_\ell(x)\tfrac12 (\nabla_i^\textnormal{S} +
 \overleftarrow\nabla_i^\textnormal{S})\psi_\textnormal{b}(x)$ & -1 &
 0.1083(12)
\end{tabular}
\end{center}
\caption{Overview of the vector current insertions in HQET at $\ord{1/\mh}$,
their corresponding tree-level matching coefficients, and the results at the
highest light-quark smearing. $\nabla^S_i$ are symmetric lattice derivatives.} 
\label{tab:vec}
\end{table}

Additional terms at $\ord{1/\mh}$ to the vector current are
\begin{align}
&V_0^{\mathrm{HQET}}(x) =
Z_{V_0}^{\mathrm{HQET}}\big(V_0^{\mathrm{stat}}(x)+{\textstyle
\sum}_{\kn=1}^2
\omega_{0,\kn}V_{0,\kn}(x)\big),\\[3pt]
&V_i^{\mathrm{HQET}}(x) =
Z_{V_i}^{\mathrm{HQET}}\big(V_i^{\mathrm{stat}}(x)+{\textstyle
\sum}_{\kn=1}^4
\omega_{i,\kn}V_{i,\kn}(x)\big)
\end{align}
where the operators used are summarized in Table~\ref{tab:vec}. Note that with
our choice of momentum, along the $x$-axis, only $i=1$ contributes. Also, for
this choice of momentum the large-volume matrix elements arising from $V_{1,3}$
and $V_{1,4}$ are identical.

\begin{figure}[tbp]
  \centering
  \includegraphics[width=11cm,clip]{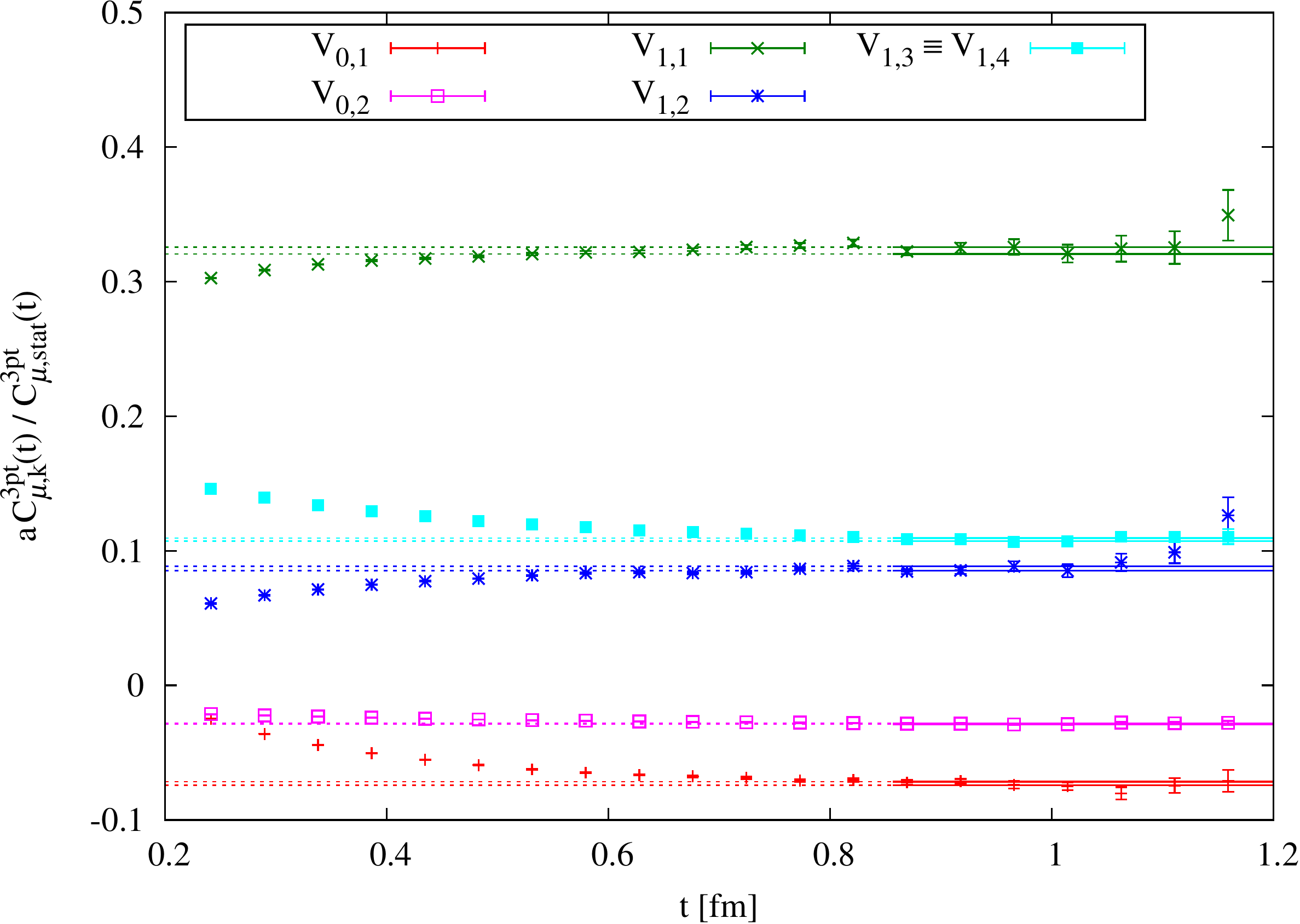}
  \caption{Overview of the current insertions. Fit bands are plateaux averages
  starting at 0.86 fm.}
  \label{fig:curr}
\end{figure}

The results obtained are presented in Fig.~\ref{fig:curr}. We see clear plateaux
starting at roughly 0.8-0.9 fm and the precision is better than that of the kin
and spin terms. A full quantitative comparison has to wait until the
corresponding non-perturbative matching coefficients are available.

\section{Summary \& outlook}

\begin{figure}[tb]
  \centering \includegraphics[width=11cm,clip]{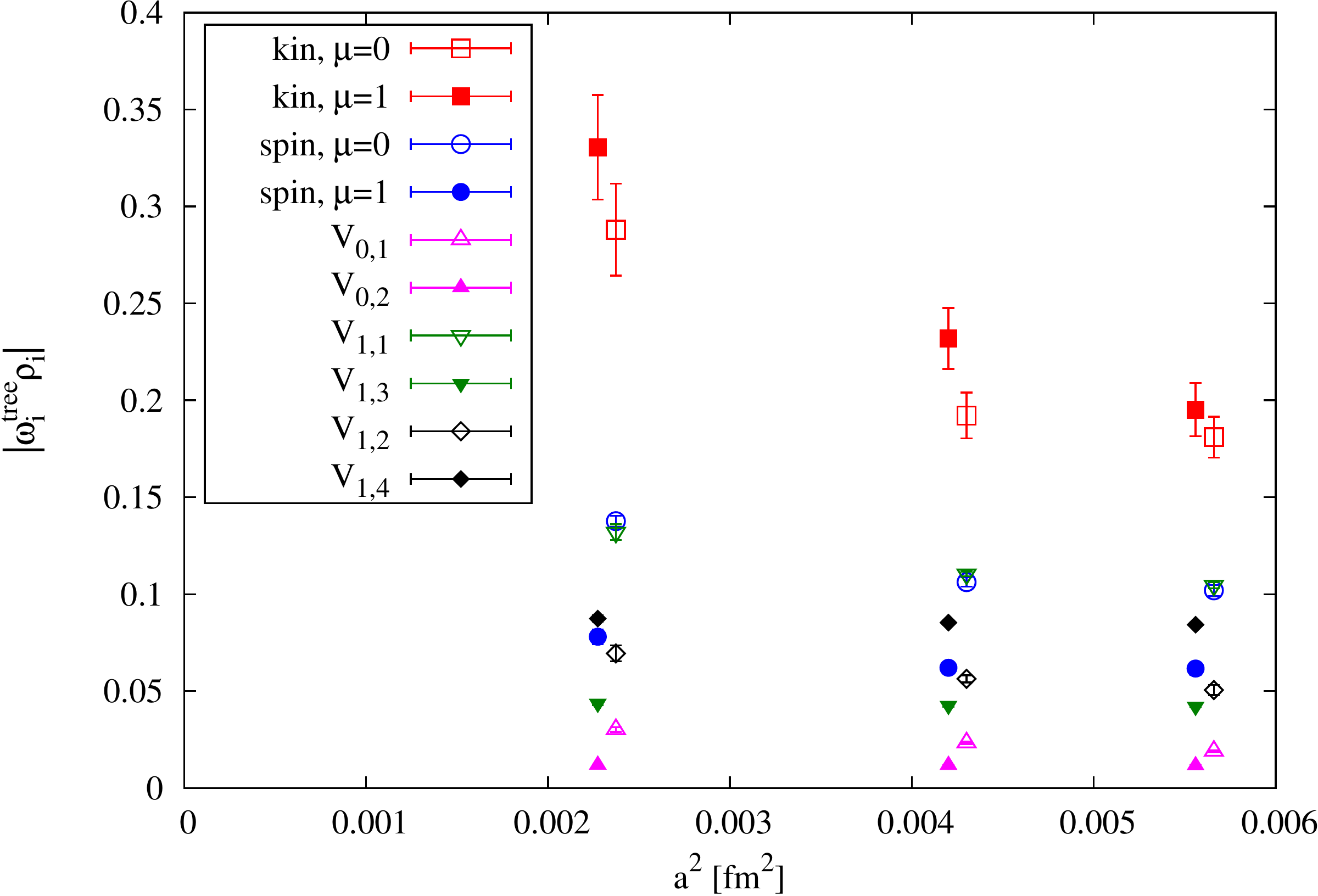}
  \caption{Absolute values of bare matrix-elements contributions, multiplied by
  their corresponding tree-level coefficients, as a function of the lattice
  spacing on CLS ensembles A5, F6 and N6 \cite{Fritzsch:2012wq}. All the
  ensembles have a similar pion mass and the momentum transfer on the coarser
  ensembles has been tuned to match that of N6 by using twisted boundary
  conditions, cf.~Ref.~\cite{Bahr:2016ayy}. Note that $1/a$ divergences are not
  removed in those bare quantities. Only after combining with the
  non-perturbative matching results, the continuum limit can be taken. The data
  for different $\mu$ are slightly shifted horizontally for greater visibility.}
  \label{fig:contlim}
\end{figure}

Let us first give a graphical summary of the obtained results. The extracted
$1/\mh$ contributions as a function of the lattice spacing on three CLS
ensembles is given in Fig~\ref{fig:contlim}. In the current form, the continuum
limit cannot be taken, as divergences will only be removed once the
non-perturbative matching coefficients are known.

The overall precision is limited by the signal-to-noise problem, the signal
rapidly deteriorates above 1.2~fm. For the kin and spin contributions the
matching coefficients are available, therefore one can give a rough estimate of
the obtained precision, which we expect to be at the level of 3\% of the final
result. This is comparable to the precision of our previously obtained static
results \cite{Bahr:2016ayy}.

For the $1/\mh$ corrections to the vector current the large-volume matrix
elements $\rho_{\mu,\kn}$ themselves are determined with an absolute precision
between $3\times 10^{-4}$ and $3\times10^{-3}$. This will give a sub-percent
contribution to the uncertainty of the final result, unless the corresponding
non-perturbative matching coefficients have unnaturally high values.

In other words, the computation of the $1/\mh$ terms is not the most challenging
part in the determination of the form factors. The same pattern for the $1/\mh$
corrections  was seen before in simpler quantities
\cite{Bernardoni:2013xba,Bernardoni:2014fva,Bernardoni:2015nqa}.

A further significant improvement of the precision in the presence of an
exponential signal-to-noise problem is not an easy task. Some gain can certainly
be obtained by using momentum smearing \cite{Bali:2016lva}. Another promising
direction is to use a multi-level algorithm along the lines of
Refs.~\cite{Ce:2016idq,Ce:2016ajy}.

\bibliography{lattice2017}

\begin{thebibliography}{15}

\bibitem{Bahr:2016ayy}
F.~Bahr, D.~Banerjee, F.~Bernardoni et~al. (ALPHA), Phys. Lett. \textbf{B757},
  473 (2016), \texttt{1601.04277}

\bibitem{Sommer:2015hea}
R.~Sommer, Nucl. Part. Phys. Proc. \textbf{261-262}, 338 (2015),
  \texttt{1501.03060}

\bibitem{Heitger:2003nj}
J.~Heitger, R.~Sommer (ALPHA), JHEP \textbf{02}, 022 (2004),
  \texttt{hep-lat/0310035}

\bibitem{DellaMorte:2013ega}
M.~Della~Morte, S.~Dooling, J.~Heitger, D.~Hesse, H.~Simma (ALPHA), JHEP
  \textbf{05}, 060 (2014), \texttt{1312.1566}

\bibitem{Fritzsch:2012wq}
P.~Fritzsch, F.~Knechtli, B.~Leder et~al., Nucl. Phys. \textbf{B865}, 397
  (2012), \texttt{1205.5380}

\bibitem{DellaMorte:2005yc}
M.~Della~Morte, A.~Shindler, R.~Sommer, JHEP \textbf{0508}, 051 (2005),
  \texttt{hep-lat/0506008}

\bibitem{Gusken:1989ad}
S.~G{\"u}sken, U.~L{\"o}w, K.H. M{\"u}tter et~al., Phys. Lett. \textbf{B227},
  266 (1989)

\bibitem{Bernardoni:2013xba}
F.~Bernardoni et~al., Phys. Lett. \textbf{B730}, 171 (2014), \texttt{1311.5498}

\bibitem{Alexandrou:1990dq}
C.~Alexandrou, F.~Jegerlehner, S.~G{\"u}sken, K.~Schilling, R.~Sommer, Phys.
  Lett. \textbf{B256}, 60 (1991)

\bibitem{Blossier:2012qu}
B.~Blossier, M.~Della~Morte, P.~Fritzsch et~al. (ALPHA), JHEP \textbf{09}, 132
  (2012), \texttt{1203.6516}

\bibitem{Bernardoni:2014fva}
F.~Bernardoni et~al. (ALPHA), Phys. Lett. \textbf{B735}, 349 (2014),
  \texttt{1404.3590}

\bibitem{Bernardoni:2015nqa}
F.~Bernardoni, B.~Blossier, J.~Bulava et~al., Phys. Rev. \textbf{D92}, 054509
  (2015), \texttt{1505.03360}

\bibitem{Bali:2016lva}
G.S. Bali, B.~Lang, B.U. Musch, A.~Schäfer, Phys. Rev. \textbf{D93}, 094515
  (2016), \texttt{1602.05525}

\bibitem{Ce:2016idq}
M.~Cè, L.~Giusti, S.~Schaefer, Phys. Rev. \textbf{D93}, 094507 (2016),
  \texttt{1601.04587}

\bibitem{Ce:2016ajy}
M.~Cè, L.~Giusti, S.~Schaefer, Phys. Rev. \textbf{D95}, 034503 (2017),
  \texttt{1609.02419}

\end{thebibliography}

\end{document}